Title: Evolution of Ion Wake Characteristics with Experimental Conditions

Authors: Rahul Banka[1], Katrina Vermillion[1], Lorin Matthews[1], Truell Hyde[1], and Lenaïc Couëdel[2,3]

*[1]Center for Astrophysics, Space Physics, and Engineering Research (CASPER), Baylor University, Waco, TX, USA*

*[2]Physics and Engineering Physics Department, University of Saskatchewan, Saskatoon, Canada*

*[3]CNRS, Aix-Marseille Université, PIIM UMR 7345, Marseille, France*

Abstract: Two-dimensional microparticle crystals can be formed in the sheath of a gas discharge plasma. Ions from the bulk plasma are accelerated in the sheath electric field, flowing past the grains to create a positive ion wake downstream from the grains. Interaction between the ion wake and neighboring grains creates additional coupling between oscillation modes and can trigger mode-coupling instability (MCI). Recent experiments have shown that at a fixed discharge power there are threshold pressures above and below which the monolayer always crystallizes or melts, respectively. The melting is due to MCI being triggered in the crystal monolayer, while the crystallization is due to the suppression of MCI by neutral damping in the fluid monolayer. The relationship between the discharge parameters and ion wake characteristics is unknown. A molecular dynamics simulation of ion dynamics and dust charging is used to self-consistently determine the dust charge and ion wake characteristics for different experimental conditions. It is found that the ion wake is strongly dependent on discharge pressure but not affected much by the discharge power.

1) Introduction:

Complex plasmas are ionized gas that include micron-sized grains. The dust grains generally become negatively charged and can self-organize into crystalline structures [1–5]. Since these dust grains are easily imaged, complex plasma crystals can be used to study events such as phase transitions at the kinetic level [6–9]. Experiments investigating phase transitions with complex plasmas are often performed in a modified GEC rf reference cell. The dust grains levitate above the lower electrode where the electrostatic force acting on the dust grains in the sheath above the lower electrode balances with gravity. While the confinement in the vertical direction is strong, the dust grains are not as strongly confined horizontally, allowing them to disperse uniformly in a 2D plane forming a hexagonal lattice. Ions from the bulk plasma are accelerated by the sheath electric field and flow past the dust grains creating a region of excess ion density downstream from the grain called the ion wake [10–15].

Ion wakes act as an independent body in the interaction between two dust grains, causing these interactions to be nonreciprocal, meaning the attractive force applied by the wake of one grain on a second grain is not necessarily equal to the attractive force applied by the wake of the second grain to the first [16]. This non-reciprocal attraction between the positive ion wake and a negatively charged neighboring dust particle can cause the in-plane (horizontal) and out-plane (vertical) oscillation modes to couple, which in turn creates an unstable hybrid mode. Thus, the energy transferred to the dust monolayer from the ions flowing past the dust grains can trigger the Mode-Coupling Instability (MCI) causing microparticles to gain energy. In a crystalline state, the energy transferred to the dust grains can lead to the breaking of symmetry and cause the structure to melt [16]. When in a fluid state, the dust particle energy (or temperature) has been observed to continue to increase, suggesting that MCIs continue to act after the crystal has melted unless suppressed by damping[17].

While the additional heating can cause a crystallized monolayer to melt, it makes it more difficult for particles in an initial fluid state to form crystalline structures. Previous experiments have shown that for microparticle monolayers levitating in the sheath of a radio-frequency discharge at a fixed discharge power there are two threshold pressures [18,19]: an upper threshold, $p_{crys}$, above which the monolayer always has a crystalline structure, and a lower threshold, $p_{MCI}$, below which the monolayer always undergoes mode-coupling instability (MCI) causing the monolayer to melt. Between these two pressures, the monolayer can be in either a crystalline or fluid state. If the monolayer is initially in a fluid state, it will remain as a fluid in the pressure range between $p_{MCI}$ and $p_{crys}$ until the pressure is increased to $p_{crys}$, at which point it will crystallize. Similarly, if the monolater is initially in a crystalline state, it will remain a crystal until the pressure is decreased to $p_{MCI}$, at which point it will become a fluid.

In a simplified model, the ion wakes can be thought of as fixed, positive point charges charge $q_w$ at a distance $l$ downstream of each dust particle. This model adequately represents the system if grains remain far enough apart that the wake charge and location relative to the dust grain are constant [15]. In previous studies, [16, 20–29], the point charge model has been used to study MCIs; however, the impact of changing discharge parameters, such as rf power and neutral gas pressure, on the ion wake parameters remains largely unknown.

The molecular dynamics simulation Dynamic Response of Ions And Dust (DRIAD) [15,30–32] are used to determine dust charge and ion wake characteristics for the different experimental conditions described in Couëdel and Nosenko article [33], hereafter referred to as Paper I. Plasma parameters that are unknown or not easily measured such as the sheath electric field, electron temperature, ion and electron number density, and ion flow speed, are determined through an iterative approach that optimizes the balance between the resultant electrostatic and gravitational forces on a dust grain for a given ion flow speed, allowing the wake characteristics to be obtained as a function of system power and pressure. Discharge parameters for stable levitating dust grains have been experimentally determined many times in the past, yet the relationship between these parameters and the characteristics of the ion wake is unknown. As such, DRIAD is utilized to model the characteristics of the ion wake for different discharge parameters.

This paper is organized as follows. Section 2 describes the experiment from which the input plasma parameters are derived. Section 3 describes the numerical model, DRIAD, and the process by which the additional plasma parameters are obtained. The results of the simulated discharge conditions are presented in Section 4 along with the calculated wake characteristics such as the total wake charge and the distance between the dust grain and the ion wake's center of charge. Section 5 is a discussion of the results and conclusions.

2) Experimental Background

This section summarizes the results from Paper I [33].

Experiments were performed in a modified GEC cell. The experimental setup is depicted in figure 1. These experiments were performed using argon gas between 0.5 and 2 Pa, and rf power between 5 and 25 W, as measured. A dust monolayer was suspended in plasma consisting of spherical melamine-formaldehyde (MF) microparticles with a diameter of $9.19 \pm 0.09 \, \mu m$.

Threshold values of $p_{MCI}$ and $p_{crys}$ were found for various rf powers (experimental procedures can be found in Paper I [33]). Two sets of experiments were performed. The first set of experiments explored power ranges from 25 W to 16 W. The second set of experiments, exploring the power range 16W to 7W, utilized the same dust monolayer as the first set of experiments but was performed about an hour later. During this time it is assumed the MF dust grains were etched, reducing their size at a rate of ~1.25 nm/min [34], implying that for the second set of experiments the dust grains were slightly smaller. The expected size difference is about 2% and is ignored in this study. The stability of a crystalline monolayer was shown to increase with rf power and argon pressure. However in Paper I, the MCI thresholds is shown to be strongly dependent on interparticle interactions and wake parameters leading to the study presented in this paper [33].

Tables I and II provide the neutral gas pressure, power, and measured effective grain charge (taken to be the excess charge on the dust grain relative to the ion wake) for the 24 experimental conditions from the first and second sets of experiments, respectively.

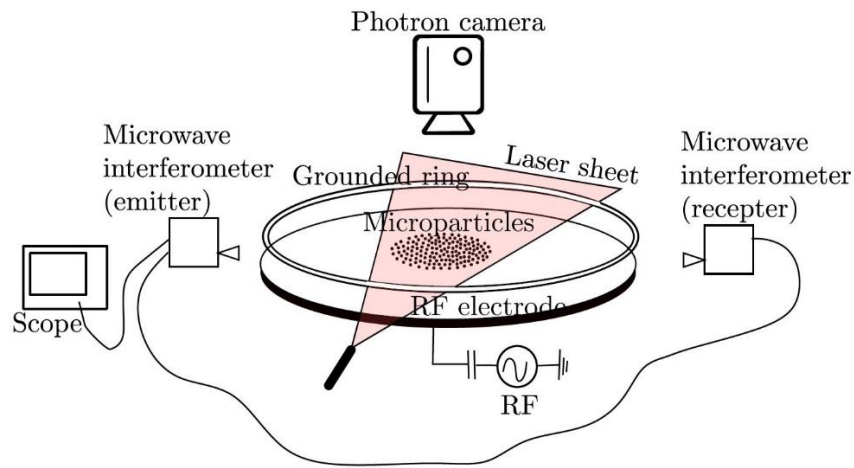

Fig.1 Schematic of the experimental setup. Reproduced from Paper I [33].

### 3) DRIAD and Use of Experimental Data for Model Inputs

The molecular dynamics simulation DRIAD was used to model the dynamics of the ions and dust and self-consistently calculate the dust charge [15,31,32]. The charge on a dust grain is determined from the electron and ion currents to the dust grain surface. The electron current is calculated using orbit motion limited (OML) theory, where the electrons are assumed to be Boltzmann distributed and are not directly modeled. The ion current to the dust grain depends on the ion flow speed as well as the dust charge and is calculated by counting the number of ions with charge $q_i$ that cross the collection radius of a dust grain, $N_{ic}$. The charge collected per time step is then $\Delta Q_{di} = N_{ic} q_i$. The ions are simulated within a cylindrical region where the z-axis is oriented along the direction of the sheath electric field. In this work, the wake region of a single, fixed dust grain was studied, where the dust grain is placed at the center of the cylindrical simulation region with a height of 10 $\lambda_{De}$ and radius of 1 $\lambda_{De}$.

Simulation parameters were based on the 24 combinations of RF power and neutral gas pressure listed in Tables I and II. The neutral gas density was calculated from the neutral gas pressure $P$ at a gas temperature $T_g = 300$ K, $n_g = P/k_B T_g$, where $k_B$ is the Boltzmann constant. The electron temperature, $T_e = 2.5$ eV, and electron density, $n_e = 2 \times 10^9$ cm$^{-3}$, are provided for a neutral gas pressure $P = 0.66$ Pa and RF power of 20 W, based upon Langmuir probe measurements performed in the same experimental setup. Using these values as a guide, further input parameters needed for the simulation were determined using a uniform density discharge model for the bulk plasma [35] to calculate the expected values for the electron temperature in each case. The electron and ion densities, $n_{e0} = n_{i0}$, were then determined based on the electron density measurements at various rf powers and pressures as reported in Figure 2 of Paper I [33]. The initial electric field for each case was determined by considering the maximum expected potential difference between the plasma bulk and the rf electrode based upon the maximum peak-to-peak rf voltage, which is a function of rf power and neutral gas pressure. These values were determined for each of the 24 cases and used as input parameters for the first iteration of DRIAD simulations. The results were checked for consistency by requiring the effective electrostatic force $(Q_d - q_w)E$ to be within 10% of the gravitational force $mg$ acting on the dust grain. Based on the results of the comparison, the electric field and ion flow velocity values were adjusted for the next iteration of DRIAD simulations. This iterative process continued until the force balance matched to within 10% and the input ion flow velocity estimated from the electric field matched the output ion flow velocity to within 13%. The electric field values determined by the iterative method correspond to the electric field determined from PIC simulations at a distance 10-14 mm above the lower electrode, as shown in Figure 2, which is consistent with experimentally observed dust levitation heights of ~10 mm [36]. The parameters for $T_e$, $n_{i0}$, and $E$ are given in Tables I and II for each of the cases.

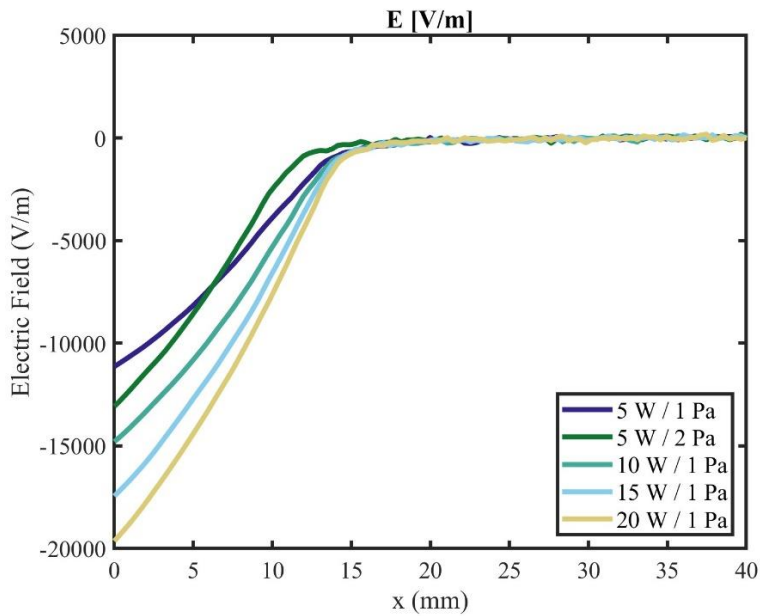

Fig 2. Results for the axial electric field determined from PIC simulations at rf powers and neutral gas pressures comparable to the conditions used in the 24 cases presented in this work.

4) Results

Each simulation was run for a physical time of 8 μs, which covers 4-8 ion plasma periods depending on the plasma conditions simulated. The plasma period, which depends on the mass of a single ion $m_i$ and the background ion density $T = 2\pi/\sqrt{((e^2\, n_{i0})/(\varepsilon_0\, m_i\,))}$. During the first 400 timesteps, corresponding to two to four ion plasma periods, the system was allowed to reach equilibrium. Results for the dust charge, ion density, and potential were then averaged over the subsequent 400 timesteps, two to four ion plasma periods. Data for the azimuthally symmetric ion density and electric potential was collected on a 2D grid in the xz-plane.

The resulting charge on the dust grain for each set of conditions is given in Tables 1 and 2. Theoretically, the surface potential of a grain with radius $a$ in a flowing plasma is given by the numerical solution of the OML electron current and the current of flowing ions [15,37,38]

$$I_e = 4\pi a^2 n_e e\left(\frac{kT_e}{2\pi m_e}\right)^{\frac{1}{2}} \exp\left(-\frac{e\Phi_d}{k_B T_e}\right), \tag{1}$$

$$I_i = \pi a^2 n_i q_i v_{dr}\left[\left(1 + \frac{1}{2\xi^2} - \frac{q_i\Phi_d}{k_B T_i \xi^2}\right)\text{erf}(\xi) + \frac{1}{\sqrt{\pi}\xi}\exp(-\xi^2)\right] \tag{2}$$

where $\xi = v_{dr}/\sqrt{2k_b T_i/m_i}$ , $n_{e,i}$, $m_{e,i}$, and $T_{e,i}$, are the electron and ion density, mass, and temperature, respectively, and $\Phi_d$ is the dust surface potential. The predicted dust charge is then

$$Q_d = 4\pi\varepsilon_0 r_d \Phi_d. \tag{3}$$

The dust charge predicted for the experimental conditions is about 1.5 times larger than the dust charge obtained in the simulation, which is consistent with the reduction in charge expected from ion-neutral collisions [39].

The ion densities in the wakes for the first and second sets of experimental conditions are shown in Fig. 3 and Fig 4, respectively. As shown, power decreases moving to the right and pressure decreases moving down. The location of the dust grain is indicated by a white circle with the center of charge indicated by a black diamond. At these very low pressures, the ion wake is greatly attenuated and stretched in the direction of the ion flow. In this work, we define the wake as the region where $n_i - n_0 > 0$. As shown in these figures, the ion density in the wake region tends to decrease as both the power and pressure decrease. The solid and dashed black lines are the equipotential lines in the region around the dust grain ranging from -0.1 to 0.15 V in 0.5 V increments, with solid lines representing negative values and dashed lines indicating positive potential.

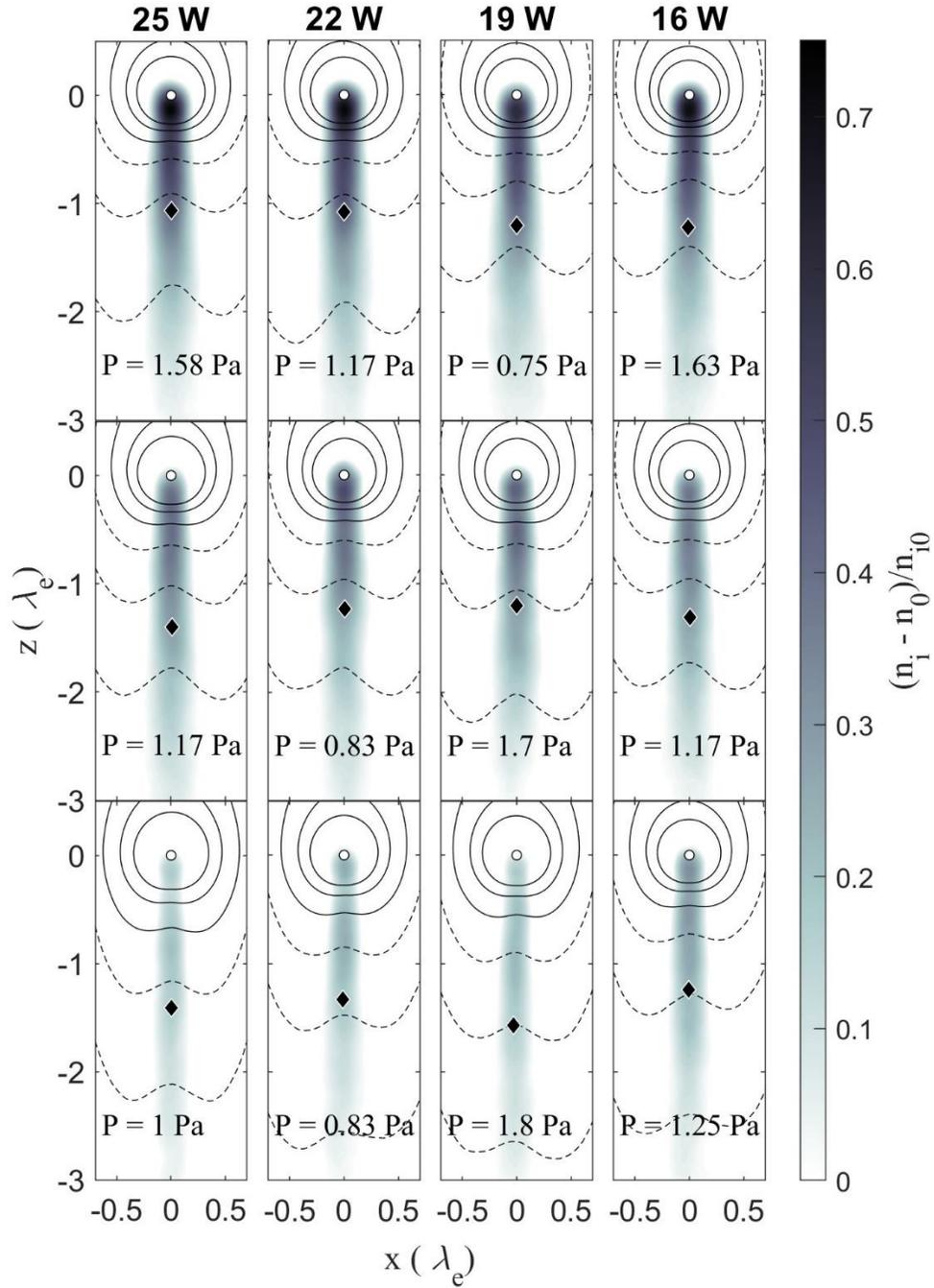

Fig. 3 Ion density in the vicinity of the dust grain for the first set of experiments. The power decreases from left to right whereas the pressure decreases from top to bottom. The grayscale image shows the wake region ($n_i - n_0)/n_0 > 0$. The contour lines indicate the potential from -0.1 to 0.15 V in steps of 0.05 V, where dashed lines are contours of positive potential and solid lines indicate negative potential contours. The black diamond is the location of the center of charge, and the white dot is the location of the center of the dust grain, size not to scale. One case with power 19 W and pressure 1.00 Pa, is not depicted in this figure.

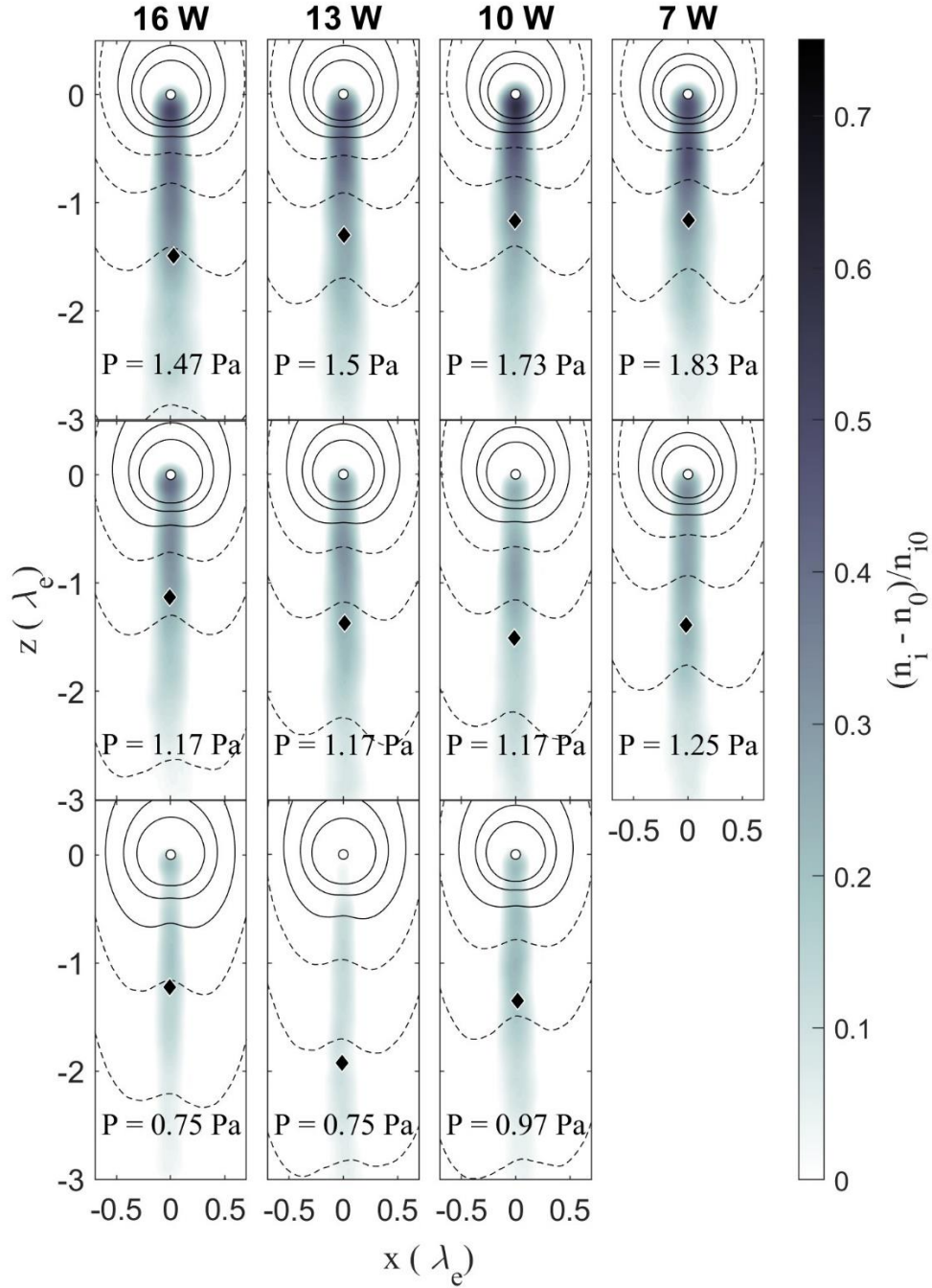

Fig 4. Ion density in the vicinity of the dust grain for the second set of experiments. The power decreases from left to right whereas the pressure decreases from top to bottom. The grayscale image shows the wake region $(n_i - n_0)/n_0 > 0$. The contour lines indicate the potential from -0.1 to 0.15 V in steps of 0.05 V, where dashed lines are contours of positive potential and solid lines indicate negative potential contours. The black diamond is the location of the center of charge, and the white dot is the location of the center of the dust grain, size not to scale.

The number of ions within the ion wake is calculated by finding the number of ions in the region $dxdz$ associated with each grid point and using azimuthal symmetry to calculate the total number of ions in a half ring at radial distance $x$ from the z-axis.

$$N_r = \pi \left( x + \frac{dx}{2} \right) (n_i - n_0) dx \, dz \qquad (4)$$

where $x$ is the horizontal distance from the dust grain to the grid location, and $dx$ and $dz$ are the grid spacings in the $x$- and $z$-directions.

The wake charge is commonly modeled by treating the ion wake as a positive point charge $q_w$, located a distance $l$ downstream of a dust grain, with the point charge assumed to be located at the center of charge of the wake region. The values describing the wake characteristics are determined by calculating the dipole moment of the wake relative to the position of the dust grain

$$\bar{p} = \sum_r q_i N_r (\bar{r_r} - \bar{r}_{dust}) \qquad (5)$$

where $q_i$ is the charge of a single ion, $\bar{r_j}$ is the location of the grid cell, and $\bar{r}_{dust}$ is the location of the dust grain. The location of the center of charge downstream of the grain is given by

$$l = \sum_r N_r \, \bar{r_r} \, / \sum_r N_r. \qquad (6)$$

The magnitude of the dipole moment is then divided by the magnitude of the distance between the dust grain and center of charge, which gives the charge in the wake. $q_w = p/l$. Although the x-components of both the dipole moment and center of charge are computed, these values are negligible since $\bar{p}_x \sim 10^{-3}$ and $l_x \sim 10^{-3} \, \lambda_{\text{de}}$.

Characteristic parameters of the wakes are plotted as a function of power and pressure in Figures 5 and 6. The electron Debye length $\lambda_{De}$ for each set of conditions is also shown. The ratio $q_w/Q_d$ and the distance to the center of charge $l$ are relatively independent of the power (Fig. 5 a,b). Note that due to the time-varying ion density and dust charge, the standard deviation of the values of $l$ and $q_w/Q_d$ is about 10%. However, as seen in Figure 6, $q_w/Q_d$ increases with the pressure while $l$ slightly decreases indicating that at higher pressures more ions are concentrated in the wake.

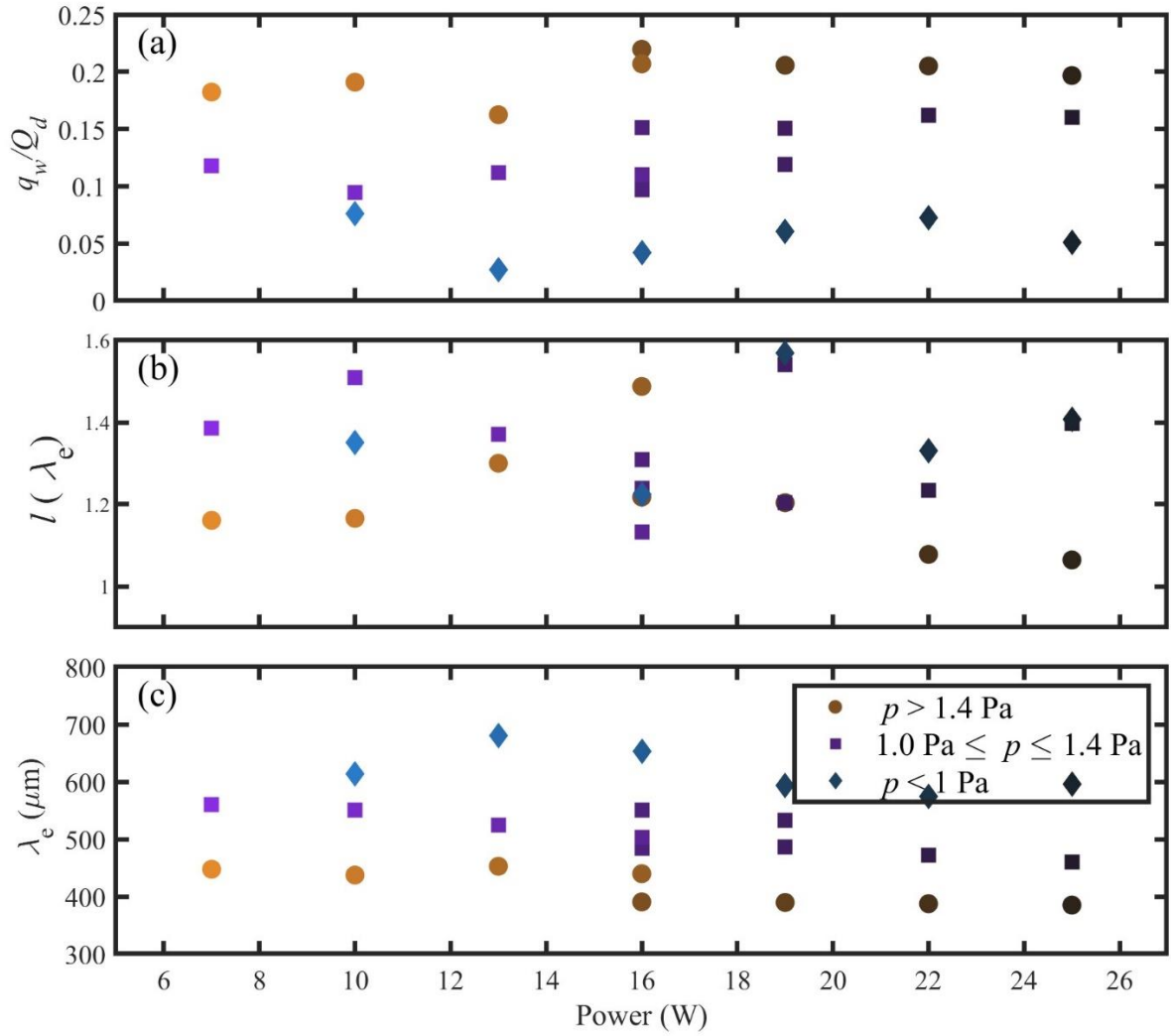

Fig. 5. Wake characteristics as a function of power. (a) The ratio between wake charge and dust charge, $q_w/Q_d$, (b) the distance between the center of charge and the dust grain $l$, and (c) the electron Debye length, $\lambda_e$. Different colors correspond to different pressure ranges. The shading of each color corresponds to different power values, with darker shades indicating higher powers.

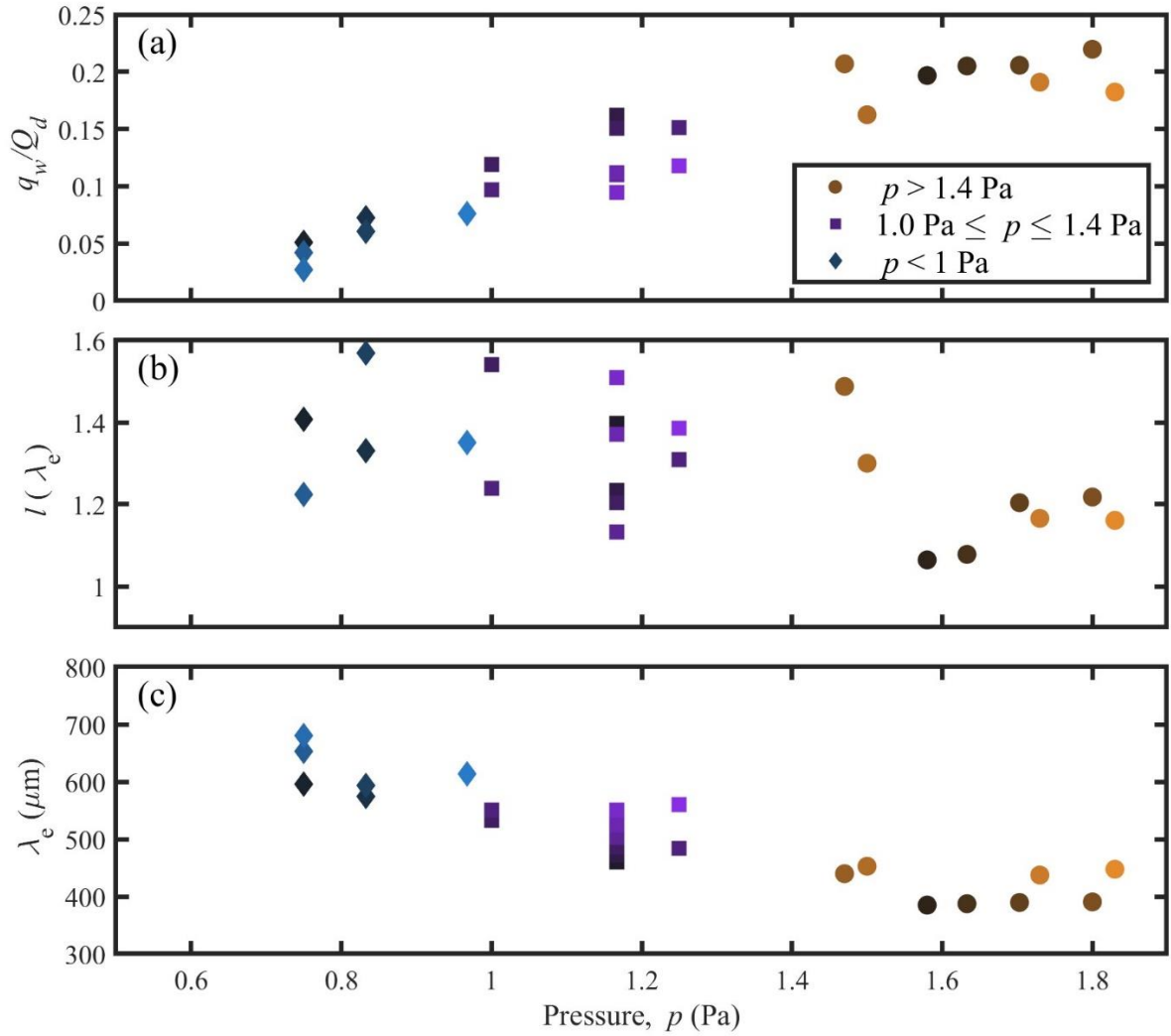

Fig. 6. Wake characteristics as a function of pressure. (a) The ratio between wake charge and dust charge, $q_w/Q_d$, (b) the distance between the center of charge and the dust grain $l$, and (c) the electron Debye length, $\lambda_e$. Different colors and symbols correspond to different pressure ranges. The shading of each color corresponds to different power levels, with darker shades indicating higher powers.

As shown in Figures 3 and 4, the potential contours at distinct z-levels directly downstream of the grain exhibit local maxima. As a general trend, the contours of positive potential move upstream towards the dust grain as the power increases. In contrast to other numerical simulations of ion wakes, the equipotential lines downstream of the dust grain do not define distinct positive potential regions relative to the background potential [15,40–42]. The primary difference in this set of simulations is the much smaller dust size, $r_{dust} \cong 5\ \mu m$, leading to weaker ion focusing. Given that the wakes in this pressure and power regime, for these small grain sizes, are very extended in the direction of the ion drift, the point charge model of the wake does not accurately capture the electrostatic potential of the system, as illustrated in Figure 7.

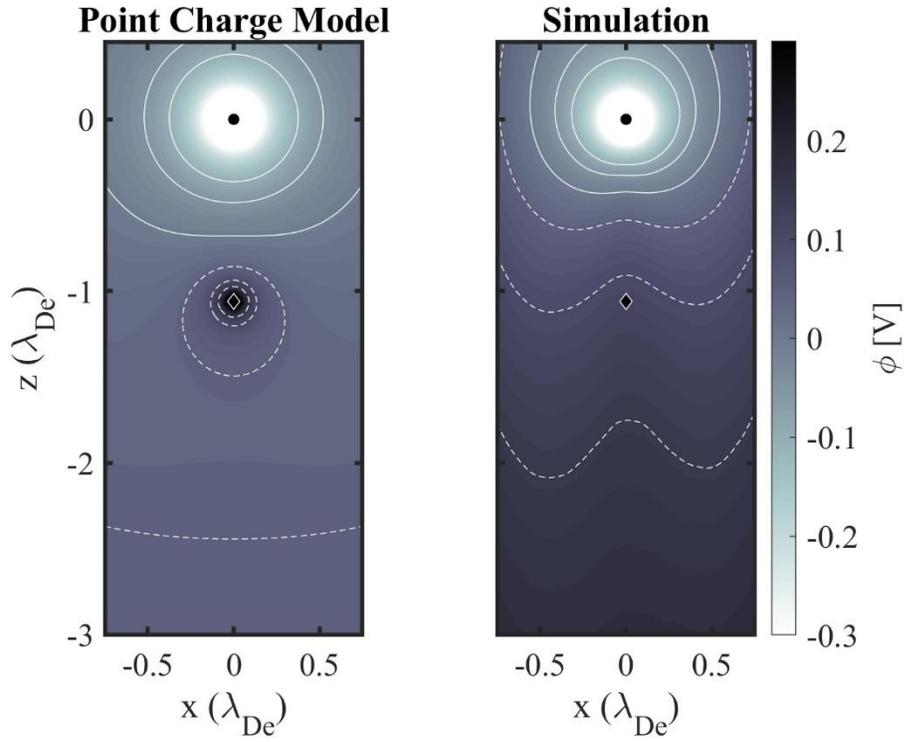

Fig 7. Contour plots of the electric potential around the dust grain. (a) The electric potential calculated by estimating the ion wake as a point charge. (b) The electric potential as outputted from the numerical simulation DRIAD. The contour lines indicate the potential from -0.1 to 0.15 V in steps of 0.05 V, where dashed lines are contours of positive potential and solid lines indicate negative potential contours. The black diamond is the location of the center of charge, and the white dot is the location of the center of the dust grains. These plots are generated for the 25W power and 1.58 Pa pressure case.

Table 1. Data associated with the first set of experiments. Experimental settings for discharge power and pressure, calculated parameters for the sheath electric field, electron temperature, ion density, electron Debye length, and ion Debye length, and simulation results for ion flow speed, dust charge, and the ion wake characteristics as described in the text.

| Experimental Conditions | | | Simulation Inputs | | | | | Simulation Results | | | |
|---|---|---|---|---|---|---|---|---|---|---|---|
| Power (W) | Pressure (Pa) | $Q_{eff}$ ($10^4$ e-) | E (V/m) | $T_e$ (eV) | $n_{i0}$ ($10^{14}$m$^{-3}$) | $\lambda_{de}$ ($\mu$m) | $\lambda_i$ ($\mu$m) | Mach | $Q_d$ ($10^4$ e-) | $l$ ($\lambda_e$) | $q_w$ (e-) |
| 25 | 1.58 | $1.05 \pm 0.29$ | 2185 | 2.15 | 8.00 | 385 | 45.5 | 1.80 | 1.79 | 1.06 | 3548 |

| | 1.17 | $1.01 \pm 0.24$ | 2100 | 2.26 | 5.91 | 460 | 53.0 | 1.65 | 1.95 | 1.40 | 3231 |
| | 0.75 | $1.46 \pm 0.30$ | 1850 | 2.44 | 3.80 | 596 | 66.1 | 2.00 | 2.16 | 1.41 | 1258 |
| 22 | 1.63 | $1.11 \pm 0.30$ | 2200 | 2.13 | 7.86 | 387 | 45.9 | 1.90 | 1.80 | 1.08 | 3708 |
| | 1.17 | $1.43 \pm 0.51$ | 2000 | 2.26 | 5.61 | 472 | 54.4 | 1.65 | 1.92 | 1.23 | 3241 |
| | 0.83 | $1.41 \pm 0.60$ | 1900 | 2.40 | 4.01 | 575 | 64.3 | 1.75 | 2.07 | 1.33 | 1660 |
| 19 | 1.70 | $1.00 \pm 0.21$ | 2350 | 2.12 | 7.72 | 390 | 46.3 | 1.60 | 1.76 | 1.20 | 3583 |
| | 1.17 | $1.17 \pm 0.23$ | 2030 | 2.26 | 5.29 | 486 | 56.0 | 1.75 | 1.95 | 1.20 | 3031 |
| | 1.00 | $1.23 \pm 0.17$ | 2035 | 2.32 | 4.53 | 532 | 60.5 | 1.70 | 2.06 | 1.54 | 2769 |
| | 0.83 | $1.21 \pm 0.28$ | 1920 | 2.40 | 3.77 | 593 | 66.3 | 1.80 | 2.10 | 1.57 | 1585 |
| 16 | 1.80 | $0.98 \pm 0.10$ | 2390 | 2.10 | 7.61 | 391 | 46.7 | 1.62 | 1.74 | 1.22 | 3801 |
| | 1.25 | $1.07 \pm 0.17$ | 2230 | 2.23 | 5.28 | 484 | 56.0 | 1.65 | 1.88 | 1.31 | 2950 |
| | 1.00 | $1.34 \pm 0.21$ | 1900 | 2.32 | 4.23 | 551 | 62.6 | 1.80 | 2.00 | 1.24 | 2110 |

Table 2. Data associated with the second set of experiments. Experimental settings for discharge power and pressure, calculated parameters for the sheath electric field, electron temperature, ion density, electron Debye length, and ion Debye length, and simulation results for ion flow speed, dust charge, and the ion wake characteristics as described in the text.

| Experimental Conditions | | | Simulation Inputs | | | | | Simulation Results | | | |
|---|---|---|---|---|---|---|---|---|---|---|---|
| Power (W) | Pressure (Pa) | $Q_{eff}$ ($10^4$ e-) | E (V/m) | $T_e$ (eV) | $n_{i0}$ ($10^{14}$ m$^{-3}$) | $\lambda_{de}$ ($\mu$m) | $\lambda_i$ ($\mu$m) | Mach | $Q_d$ ($10^4$ e-) | $l$ ($\lambda_e$) | q (e-) |
| 16 | 1.47 | $1.09 \pm 0.11$ | 2250 | 2.17 | 6.21 | 440 | 51.7 | 1.55 | 1.82 | 1.49 | 3847 |
| | 1.17 | $1.21 \pm 0.21$ | 2030 | 2.26 | 4.93 | 503 | 58.0 | 1.95 | 1.93 | 1.13 | 2185 |
| | 0.75 | $1.31 \pm 0.32$ | 1790 | 2.44 | 3.17 | 653 | 72.3 | 2.05 | 2.14 | 1.22 | 1046 |
| 13 | 1.50 | $1.04 \pm 0.19$ | 2250 | 2.17 | 5.84 | 453 | 53.3 | 1.70 | 1.84 | 1.30 | 3068 |
| | 1.17 | $1.04 \pm 0.12$ | 2100 | 2.26 | 4.54 | 525 | 60.4 | 1.80 | 1.98 | 1.37 | 2397 |
| | 0.75 | $1.42 \pm 0.37$ | 1800 | 2.44 | 2.92 | 680 | 75.4 | 1.80 | 2.13 | 1.93 | 1000 |
| 10 | 1.73 | $1.06 \pm 0.19$ | 2260 | 2.11 | 6.11 | 437 | 52.1 | 1.65 | 1.74 | 1.17 | 3369 |
| | 1.17 | $1.13 \pm 0.17$ | 2240 | 2.26 | 4.12 | 551 | 63.4 | 1.75 | 1.96 | 1.51 | 2101 |
| | 0.97 | $1.31 \pm 0.37$ | 1825 | 2.33 | 3.42 | 614 | 69.6 | 1.95 | 2.02 | 1.35 | 1749 |
| 7 | 1.83 | $1.14 \pm 0.20$ | 2250 | 2.09 | 5.76 | 448 | 53.7 | 1.80 | 1.74 | 1.16 | 3210 |
| | 1.25 | $1.16 \pm 0.12$ | 2240 | 2.24 | 3.93 | 561 | 65.0 | 1.65 | 1.88 | 1.39 | 2475 |

## 5) Conclusions

Utilizing a numerical model of ion dynamics, the equilibrium dust charge and ion wake characteristics are determined for varying experimental conditions. The effective dust charge, the sum of the wake charge and the dust charge, was required to be consistent with experimental measurements within $\pm 10\%$. An iterative method was used to determine the sheath electric field which both balanced the gravitational force and provided the ion flow, which determines the dust charge. The electric field found by this approach is consistent with the electric field at the expected levitation height calculated with PIC simulations of the rf discharge. The simulation results include the dust charge, ion density, and electrostatic potential in the region surrounding the grain for 24 combinations of rf power and neutral gas pressure. These results were utilized to calculate ion wake characteristics such as the equivalent wake charge for a point charge model, $q_w$, and distance between the dust grain and the center of charge, $l$.

As shown in figure 6, the charge contained in the wake $q_w/Q_d$, is proportional to the neutral gas pressure, ranging from 5% of the dust charge at pressures less than 1 Pa to 20% of the dust charge at pressures of 1.8 Pa. The charge contained in the wake is largely independent of rf power as shown in figure 5. The normalized distance between the dust grain and the location of the point wake charge varies little with power and pressure, and for all the conditions presented here is in the range of $1 < l/\lambda_{De} < 1.6$ .

   Transitions between states, as caused by MCIs, can be studied utilizing N-body simulations of a cloud of dust grains, with wake charges $q_w$ located a distance $l$ downstream from each dust grain.  Simulations where both parameters are allowed to evolve as rf power and neutral gas pressure change can be used to further understand these non-reciprocal interactions and their effects on the system dynamics. While the wake characteristics as presented above will allow for better modeling of MCIs, the point charge model may not be suitable for describing ion wakes. In addition, the wake structure will evolve in the proximity of a second grain.  A model that treats ion wakes as extended clouds of positive charge with location and magnitude depending on power, pressure, and the positive relative to close grains would more closely capture wake characteristics and allow utility across a larger range of discharge parameters. This is a subject ripe for development using machine-learning techniques to characterize the wakes.